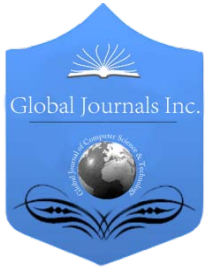

# Vertical Handover decision schemes using SAW and WPM for Network selection in Heterogeneous Wireless Networks

By K.Savitha, Dr.C.Chandrasekar

*Periyar University*

*Abstract-* Seamless continuity is the main goal and challenge in fourth generation Wireless networks (FGWNs), to achieve seamless connectivity "HANDOVER" technique is used, Handover mechanism are mainly used when a mobile terminal(MT) is in overlapping area for service continuity. In Heterogeneous wireless networks main challenge is continual connection among the different networks like WiFi, WiMax, WLAN, WPAN etc. In this paper, Vertical handover decision schemes are compared, Simple Additive Weighting method (SAW) and Weighted product model (WPM) are used to choose the best network from the available Visitor networks (VTs) for the continuous connection by the mobile terminal. In our work we mainly concentrated to the handover decision phase and to reduce the processing delay in the period of handover. In this paper both SAW and WPM methods are compared with the Qos parameters of the mobile terminal (MT) to connect with the best network.

*Keywords:* Handover, Vertical handover decision schemes, Simple additive weighting, Weight product method.

*GJCST Classification:* C.2.1

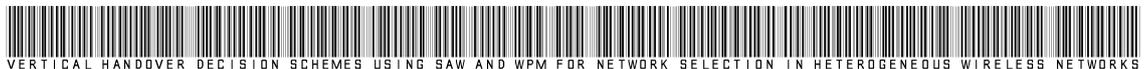

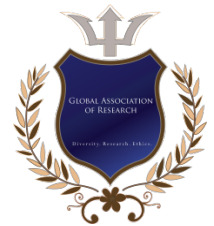

# Vertical Handover decision schemes using SAW and WPM for Network selection in Heterogeneous Wireless Networks

K.Savitha[α], Dr.C.Chandrasekar[Ω]

*Abstract-* Seamless continuity is the main goal and challenge in fourth generation Wireless networks (FGWNs), to achieve seamless connectivity "HANDOVER" technique is used, Handover mechanism are mainly used when a mobile terminal(MT) is in overlapping area for service continuity. In Heterogeneous wireless networks main challenge is continual connection among the different networks like WiFi, WiMax, WLAN, WPAN etc. In this paper, Vertical handover decision schemes are compared, Simple Additive Weighting method (SAW) and Weighted product model (WPM) are used to choose the best network from the available Visitor networks (VTs) for the continuous connection by the mobile terminal. In our work we mainly concentrated to the handover decision phase and to reduce the processing delay in the period of handover. In this paper both SAW and WPM methods are compared with the Qos parameters of the mobile terminal (MT) to connect with the best network.

*Keyword-* Handover, Vertical handover decision schemes, Simple additive weighting, Weight product method.

## I. Introduction

In fourth generation wireless networks service continuity is a main goal ie., when a MT or mobile node (MN) moving in an overlapping area, continuous service must be need so the technique "HANDOVER" is done. The handover technique is mainly used to redirect the mobile user's service network from current network to a new network or one base station (BS) to another BS or one access point (AP) to another AP with same technology or among different technologies to reduce the processing delay in the overlapping area.

*About[α]-* The Master of Science from the Bharathiar University, India in 2006, M.Phil Degree from periyar university, India in 2007. She is a Research Scholar in Department of computer science , Periyar University, Salem, India. She is pursing her Ph.D in Mobile computing. Her research area interest includes Networking, Multimedia.
(Email: ksavi_tha@yahoo.com)

*About[β]-* Ph.D. degree from Periyar University, Salem. He has been working as Associate Professor at Dept. of Computer Science, Periyar University, Salem – 636 011, Tamil Nadu, India. His research interest includes Wireless networking, Mobile computing, Computer Communication and Networks. He was a Research guide at various universities in India. He has been published more than 50 technical papers at various National/ International Conference and Journals.
(Email: ccsekar@gmail.com)

Handover technique has the two types, horizontal handover and vertical handover. The homogenous wireless network performs horizontal handover, if there are two BSs using the same access technology, in current system called horizontal handover. This type of mechanism use signal strength measurements for surrounding BSs to trigger and to perform the handover decision.

In heterogeneous wireless networks, the mobile station (MS) or BS will be equipped with multiple network interfaces to reach different wireless networks. When an emerging mix of overlapping heterogeneous wireless networks deployed, vertical handover is used among the networks using different access technologies.

Handover technique has the four phases: Handover Initiation, System discovery, Handover decision, Handoff execution.

- Handoff Initiation phase : The handover process was modified by some criteria value like signal strength, link quality etc.,
- System discovery phase: It is used to decide which mobile user discovers its neighbour network and exchanges information about Quality of Service (QOS) offered by these networks.
- Handover Decision phase: This phase compares the neighbour network QOS and the mobile users QOS with this QOS decision maker makes the decision to which network the mobile user has to direct the connection.
- Handoff Execution phase: This phase is responsible for establishing the connection and release the connections and as well as the invocation of security service.

The scope of our work is mainly in handover decision phase, as mentioned in the decision phase; decision makers must choose the best network from available networks. In this paper, the decision makers are Simple additive weighting (SAW) and Weighted product model (WPM) to take the decision and to select the best target visitor network (TVN) from several visitors networks.

In this paper, two vertical handover decision schemes (VHDS), Distributed handover decision scheme (DVHD) and Trusted Distributed vertical handover decision schemes (T-DVHD) are used. DVHD is advanced than the centralised vertical handover decision scheme and T-DVHD is the extended work of DVHD. Here we compare the distributed and trusted vertical handover decision schemes as distributed decision tasks among networks to decrease the processing delay caused by exchanging information messages between mobile terminal and neighbour networks. To distribute the decision task, vertical handover decision is formulated as MADM problem.

In our work, the proposed decision making method use WPM in a distributed manner and compared with SAW method. The bandwidth, delay, jitter and cost are the parameters took by the MT as the decision parameters for handover.

## II. Related Work

At present many of the handoff decision algorithms are proposed in the literature. In (4) a comparison done among SAW, Technique for Order Preference by Similarity to Ideal Solution(TOPSIS), Grey Relational Analysis (GRA) and Multiplicative Exponent Weighting (MEW) for vertical handoff decision. In (3) author discuss that the vertical handoff decision algorithm for heterogeneous wireless network, here the problem is formulated as Markov decision process. In (5) the vertical handoff decision is formulated as fuzzy multiple attribute decision making (MADM).

In (8) their goal is to reduce the overload and the processing delay in the mobile terminal so they proposed novel vertical handoff decision scheme to avoid the processing delay and power consumption. In (7) a vertical handoff decision scheme DVHD uses the MADM method to avoid the processing delay. In (10) the paper is mainly used to decrease the processing delay and to make a trust handoff decision in a heterogeneous wireless environment using T-DVHD.

In (11) a novel distributed vertical handoff decision scheme using the SAW method with a distributed manner to avoid the drawbacks. In (14) the paper provides the four steps integrated strategy for MADM based network selection to solve the problem. All these proposal works are mainly focused on the handoff decision and calculate the handoff decision criteria on the mobile terminal side and the discussed scheme are used to reduce the processing delay by the calculation process using MADM in a distributed manner.

In (16) the comparison analysis shows the SAW, MEW, TOPSIS, VIKOR, GRA and WMC with the numerical simulation of vertical handoff in 4G networks.

## III. Vertical Handover Decision Schemes

Centralized vertical handover decision (C-VHD), Distributed vertical handover decision (D-VHD), Trusted Distributed vertical handover decision (T-DVHD) are the schemes used to reduce the processing delay between the mobile node and neighbour network while exchanging the information during the handover. In this paper, D-VHD and T-DVHD schemes are compared. MADM have several methods in literature [16]. TOPSIS is used in distributed manner for network selection.

a) *Centralized vertical handover decision Schemes*

In C-VHD, a Mobile Node (MN) exchanging the information message to the Neighbour networks mean processing delay was increased by distributing in centralized manner. When processing delay had increased overall handover delay increases. This is one of main disadvantage in C-DHD, so Distributed Vertical handover decision (D-VHD) schemes was proposed in [7][8].

b) *Distributed vertical handover decision schemes*

D-VHD is used to decrease the processing delay than the C-VHD schemes. This scheme is mainly used for handover calculation to the Target visitor networks (TVNs). TVN is the network to which the mobile node may connect after the handover process was finished. In our work D-VHD takes into account : jitter, cost, bandwidth, delay as evaluation metrics to select a suitable VN which applied in MADM method.

c) *Network Selection Function (NSF):*

The network selection decision process has denoted as MADM problem, NSF have used to evaluate from set of network using multiple criteria. The above mentioned parameters are used to calculate NSF. These parameters measure the Network Quality Value (NQV) of each TVN. The highest NQV value of TVN will be selected as Visited Network (VN) by the mobile node. The generic NSF is defined by using SAW "Eq. (1) and WPM "Eq. (2)"

$$NQV_i = \sum_{i=1, j=1}^{N, n_p^+} W_j * P_{ij} \qquad (1)$$

Where, NQV$_i$ represents the quality of i$^{th}$ TVN. W$_j$ is the weight of the P$_{ij}$, P$_{ij}$ represents the j$^{th}$ parameter of the i$^{th}$ TVN. N is the number of TVNs. While n$_p^+$ is the number of parameters.

$$NQV_i = R_i = \frac{\prod_{j=1}^{n} x_{ij}^{w_j}}{\prod_{j=1}^{n} (x_j^*)^{w_j}} \qquad (2)$$

Where, NQV$_i$ represents the quality of i$^{th}$ TVN. $w_j$ is the weight of the attribute values, $x_{ij}$ is the positive attributes and $x_j^*$ is the negative attribute. $R_i$ is the value ratio between network I and positive ideal.

Based on the user service profile, handover decision parameters have assigns different "Weights" to determine the level of importance of each parameter. In equation (3), the sum of these weights must be equal to one.

$$\sum_{j=1}^{n_p} W_j = 1 \qquad (3)$$

The handover decision metrics calculation is performed on the VNs, each VN applies the MADM methods using "Eq. (3.1,3.2)" on the required ($J_{req}$, $D_{req}$, $C_{req}$, $B_{req}$) and offered ($J_{off}$, $D_{off}$, $C_{off}$, $B_{req}$) parameters

d)  *Distributed Decision scheme:*
The D-VHD is explained in the Fig. 3

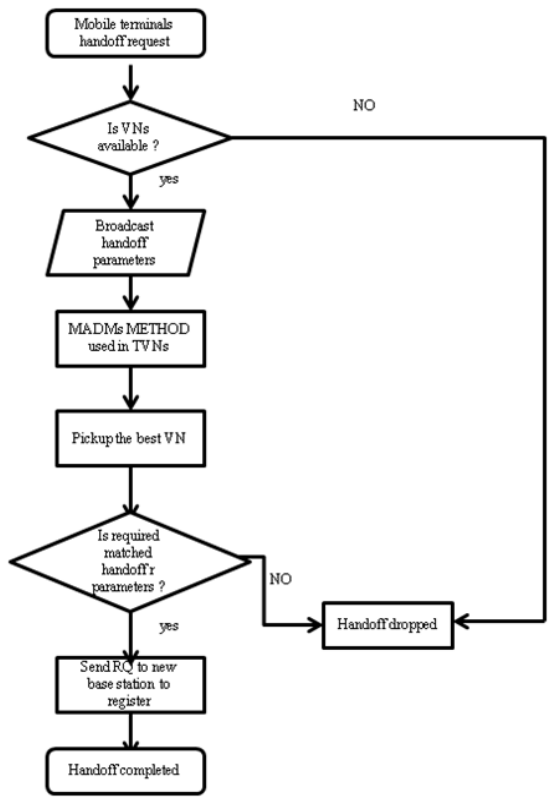

*Fig. 1* D-VHD Scheme

e)  *Trusted Distributed Vertical Handover Decision schemes*

Trusted handover decision and to avoid the unnecessary handover events are the important factors while exchanging the trusted information between networks and mobile node. The extension work of the DVHD scheme is T-DVHD scheme. The scheme is mainly introduced [10] for decreasing the processing delay than DVHD scheme.

The T-DVHD schemes followed by the DVHD Network selection function and Distribute Decision schemes, before sending request to connect a new base station trusted process is started

f)  *Level Of Trust (LOT) test function*
LOT function is tested to execute the handover. LOT function is calculated by the following steps
    If $LoT_i >=$ threshold
        Connect to the $TVN_i$
        start Trust-test function
    else if $LoT_i <$ threshold {
        if (suitable-TVN available)
            i = i + 1
            test another network
        else if (no suitable-TVN)
    Handover blocked

after handover is executed by the mobile terminal with the proper TVN. Trusted Test Function is started, once the mobile terminal connects to the TVN trusted test function is calculated by the following steps to finish the T-DVHD schemes.
if $Q_{off} < Q_{req}$
    $LOT_i = LOT - delta$ ;
 else
    $LOT_i = LOT_i + delta^+$ ;

## IV. SCENARIO OF THE VERTICAL HANDOVER

In this paper, our scenario was in "Fig. 4", it explains that a cell coverage the area by WiMax technology and another cell coverage the area by WiFi and WiMax technology. A mobile terminal is overlapping with VoIP application between the cell coverage now mobile terminal intend to connect the appropriate visited network with the decision process.

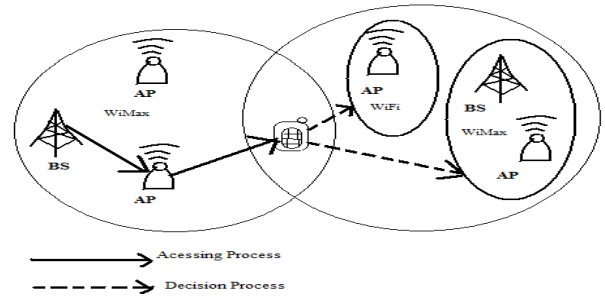

*Fig. 2* Scenario of the vertical handover

## V. SAW AND WPM

a)  *Simple Additive Weighting (SAW) Method*

Simple Additive Weighting (SAW) which is also referred as weighted linear combination or scoring methods or weighted sum method is a simple and most often used multi attribute decision technique. The method is based on the weighted average. An evaluation score is calculated for each alternative by multiplying the scaled value has given to the alternative of that attribute with the weights of relative importance directly assigned by decision maker followed by summing of the products for all criteria.

The application of SAW scoring requires, identification of objectives and alternatives, evaluation of alternatives, determination of sub-objective weights, additive aggregation of weighted partial preference values, sensitive analysis. It uses direct rating on the standardised scales only in purely qualitative attributes. For numerical attributes score are calculated by normalized values to match the standardised scale. The SAW is a comparable scale for all elements in the decision matrix, the comparable scale obtained by $r_{ij}$ for benefit criteria "Eq. (4)" and worst criteria "Eq.(5)".

$$V_{ij} = \frac{x_{ij}}{x_j^{max}} \quad (4)$$

$$V_{ij} = \frac{x_j^{min}}{x_j} \quad (5)$$

The SAW method, underlying additive values function and compute as alternatives score
$V_i = V(A_i)$ by adding weighting normalized values $W_j V_{ij}$ ∀ j = {1,………m} before eventually ranking alternatives

$$V_i = \sum_{j=1}^{m} W_j V_{ij} \quad (6)$$

For V ε $R^{n*m}$ with i = { 1,……,n}, j = {1,……..,m}; $V_{ij}$, $W_j$ ε (0,1)

b) *Weighted Product Model (WPM) method*

The weighted product model (WPM) similar to the weighted sum model (WSM) and it is also called as Multiplicative exponent Weighting (MEW). It is another MADM scoring method. The main difference is that instead of addition usually mathematical operation now there is multiplication. As with all MADM methods, WPM is a finite set of decision alternatives described in terms of several decision criteria. The vertical handover decision problem can be expressed as a matrix form and each row i corresponds to the candidate network i and each column j corresponds to the attributes.

$$V(A_i) = \prod_{j=1}^{n} x_{ij}^{w_j} \quad (7)$$

Where $x_{ij}$ denotes attribute j of candidate network i, $w_j$ denotes the weight of attributed j.

Note that in eqn. (7), $w_j$ is a positive power for benefit metrics $x_{ij}^{w_j}$, and a negative power for cost metrics $x_{ij}^{-w_j}$. Since the score of a network obtained by MEW does not have an upper bound, it is convenient for comparing each network with the score of the positive ideal network .This network is defined as the network with the best values in each metric. For a benefit metric, the best value is the largest. For a cost metric, the best value is the lowest.

$$R_i = \frac{V(A_i)}{V(A^*)} = \frac{\prod_{j=1}^{n} x_{ij}^{w_j}}{\prod_{j=1}^{n} (x_j^*)^{w_j}} \quad (8)$$

c) *Numerical Example*

The above section outlines the vertical handover decision schemes and MADM methods, SAW and TOPSIS which is used for the network selection in this paper. For instance, suppose a mobile terminal is currently connected to a WiFi cell and has to make decision among six candidate networks A1, A2, A3, A4, A5, A6, where A3, A4 are WiFi cells and others are WiMax cells. Vertical handover criteria considered here are delay, bandwidth, cost, jitter which denoted as X1, X2, X3, X4 respectively. Decision matrix D is as follows

D =

|    | X1    | X2    | X3    | X4    |
|----|-------|-------|-------|-------|
| A1 | 0.984 | 0.533 | 0.667 | 0.438 |
| A2 | 1     | 0.1   | 0.75  | 0.812 |
| A3 | 0.984 | 1     | 0.5   | 0.061 |
| A4 | 1     | 0.467 | 1     | 1     |
| A5 | 0.984 | 0.733 | 0.6   | 0.119 |
| A6 | 0.968 | 0.667 | 0.667 | 0.263 |

The users running application was voice. The preference on handover criteria is modeled as weights assigned by the user on the criteria, for voice $W_v$ which shown in the "Eq. (9)".

$$W_v = [0.3\ 0.2\ 0.2\ 0.3] \quad (9)$$

MADM methods handle in this paper for decision problems with above data. The following section discussed the SAW and WPM are applied and the results are compared.

i. *SAW*

SAW requires a comparable scale for all elements in the decision matrix, the comparable scale is obtained by using "Eq. (4), Eq. (5)". In these $x_{ij}$ is the performance score of alternatives $A_i$ with respect to criteria $x_j$. After scaling, the normalized decision matrix is evaluated as D'

D' =

|    | X1      | X2   | X3 | X4    |
|----|---------|------|----|-------|
| A1 | 0.00062 | 8    | 9  | 0.411 |
| A2 | 0.00063 | 1.5  | 8  | 0.762 |
| A3 | 0.00062 | 15   | 12 | 0.057 |
| A4 | 0.00063 | 7    | 6  | 0.939 |
| A5 | 0.00062 | 11   | 10 | 0.103 |
| A6 | 0.00061 | 1    | 9  | 0.247 |

Applying the weight factor from the "Eq. (9)", weighted average values for A1, A2, A3, A4, A5 and A6 are calculated for the respected to the voice application $A_v$

$$A_V = [0.664, 0.714, 0.563, 0.793, 0.595, 0.635]$$

The best network is A4 which is the network selected to connect the mobile terminal for service continuity with the minimum processing delay.

ii. *WPM*

The WPM is called dimensionless analysis because its mathematical structure eliminates any units of measure. Transformation is not necessary when we use multiplication among attribute values. The weights become exponents associated with each attribute values. From "Eq. (9)" the weight factor is applying for $V(A_i)$

$$V(A_i) = [0.054, 0.065, 0.024, 0.065, 0.035, 0.042]$$

$$V(A_i) = 0.074$$

$$R_i = [0.73, 0.89, 0.32, 0.88, 0.47, 0.57]$$

iii. *Ranking Order for Saw and WPM methods*

The ranking order using different methods of MADM are summarised in "Table 1". SAW and WPM ranks A4 and So the A4 and A2 BS have connected the mobile terminal with less processing delay to get seamless handover in between the MT and BS A4,A2 in each method.

*Table 1: Ranking order comparison*

| SAW | A4 | A2 | A1 | A6 | A5 | A3 |
|---|---|---|---|---|---|---|
| WPM | A2 | A4 | A1 | A6 | A5 | A3 |

After Ranking the BS for each method, we go to compare the SAW and WPM using Relative Standard Deviation (RSD).

$$RSD = \frac{S}{\bar{x}} * 100$$

The relative standard deviation is often times more convenient. It is expressed in percent and is obtained by multiplying the standard deviation s by 100 and Divide this product by the average $\bar{x}$

We calculate the relative standard deviation for SAW and MEW from ranking values. The result was WPM is better than SAW. SAW with the 12.64% and WPM with the 35.75%. Finally MT is connected to the base station with the WiMax cell A2.

## VI. Conclusion

In this paper, we have compared the schemes of vertical handover decision in the heterogeneous wireless networks. The observation of schemes to reduce the processing delay and a trusted handover decision is done in heterogeneous wireless networks. In this paper we proposed decision makers SAW and WPM to select the best network from the visitor network for the Vertical decision schemes. The best decision maker is analyzed by the relative standard deviation and the best one is WPM. Our main goal is in the decision phase of the handover phases to take decision to which VN the mobile terminal to connect to decrease the processing delay by different decision algorithms.

## References Références Referencias